\documentclass[aps,prl,english,twocolumn,showpacs,10pt,superscriptaddress]{revtex4-1} \pdfoutput=1
\usepackage{amsmath,bm,amsfonts}
\usepackage{mathtools}
\usepackage{commath}
\usepackage{amssymb}
\usepackage{graphicx}
\usepackage{babel}
\usepackage{cases}
\usepackage[colorlinks=true, citecolor=blue, anchorcolor=green]{hyperref}
\hypersetup{linkcolor=red}
\usepackage{natbib}
\usepackage{floatrow}
\usepackage{dblfloatfix}
\usepackage{xcolor}
\usepackage{braket}

\newcommand{\Rmnum}[1]{\expandafter\@slowromancap\romannumeral #1@}

\newcommand{\fsr}{\Delta_{\text{fsr}}}
\newcommand{\tgb}{\tilde{g}_B}
\newcommand{\tbeta}{\tilde{\beta}}

\newcommand{\be}{\begin{equation}}
\newcommand{\ee}{\end{equation}}
\newcommand{\bg}{\begin{aligned}}
\newcommand{\eg}{\end{aligned}}

\makeatother

\begin{document}
\title{ Quantum State Transfer via a Multimode Resonator}
\author{Yang He}
\affiliation{Institute of Physics, Chinese Academy of Sciences, Beijing 100190, China}
\affiliation{School of Physical Sciences, University of Chinese Academy of Sciences, Beijing 100049, China}
\author{Yu-Xiang Zhang}
\email{iyxz@iphy.ac.cn}
\affiliation{Institute of Physics, Chinese Academy of Sciences, Beijing 100190, China}
\affiliation{School of Physical Sciences, University of Chinese Academy of Sciences, Beijing 100049, China}
\affiliation{Hefei National Laboratory, Hefei, 230088, China}

\date{\today}

\begin{abstract}
Large-scale fault-tolerant superconducting quantum computation needs
rapid quantum communication to network qubits fabricated on different chips  and
long-range couplers to implement efficient quantum error-correction codes.
Quantum channels used for these purposes are best modeled by multimode resonators, which lie between single-mode cavities and waveguides with a continuum of modes.
In this Letter, we propose a formalism for
quantum state transfer using
coupling strengths comparable to the channel's free spectral range ($g{\sim}\fsr$).
Our scheme merges features of both the STIRAP-based methods for single-mode cavities and 
the pitch-and-catch protocol for long waveguides, integrating their advantage of 
low loss and high speed. It is immune to thermal channel occupations if using harmonic resonators
for the sender and receiver.
\end{abstract}

\maketitle

Practical quantum computation may require millions of 
qubits~\cite{Fowler:2012aa}, 
while state-of-art technology holds only
hundreds of qubits on a single superconducting processor ~\cite{Castelvecchi:2023aa}. Thus, 
it is necessary to network
many processors by 
shuttling quantum information and distributing quantum entanglement~\cite{Bravyi:2022aa} 
using schemes of quantum state transfer (QST), 
in the same manner of
quantum internet~\cite{Kimble:2008aa,Wehner:2018aa}.
To connect processors kept in the same or nearby
dilution fridges, we expect the quantum links to range from 
centimeters to a few meters. 
We refer to this regime as \emph{medium range}, in order to
distinguish it from the cases of extremely
short or long distances. Light-matter interaction
in this regime is also relevant to 
long-range couplers that are necessary for quantum low-density-parity-check
error-correction codes to significantly save the 
overhead of physical qubits~\cite{Baspin:2022aa,Bravyi:2024aa}.
Although QST has been extensively 
studied~\cite{Cirac:1997aa,Christandl:2004aa,Yao:2005aa,Kay:2010aa,
Korotkov:2011aa,Steffen:2013aa,Wenner:2014aa,Reiserer:2015aa,Xiang:2017aa,Vermersch:2017aa,
Axline:2018aa,Kurpiers:2018aa,Wang:2012aa,Magnard:2020aa,Burkhart:2021aa,Bienfait:2019aa,Zhong:2021aa,Niu:2023aa,Grebel:2024aa,McKay:2015aa,Sundaresan:2015aa,Chakram:2022aa,
Chang:2020aa}, 
the medium-range QST needs to be revisited,
for the following reasons.

\begin{figure}[b]
    \centering
    \includegraphics[width=0.85\textwidth]{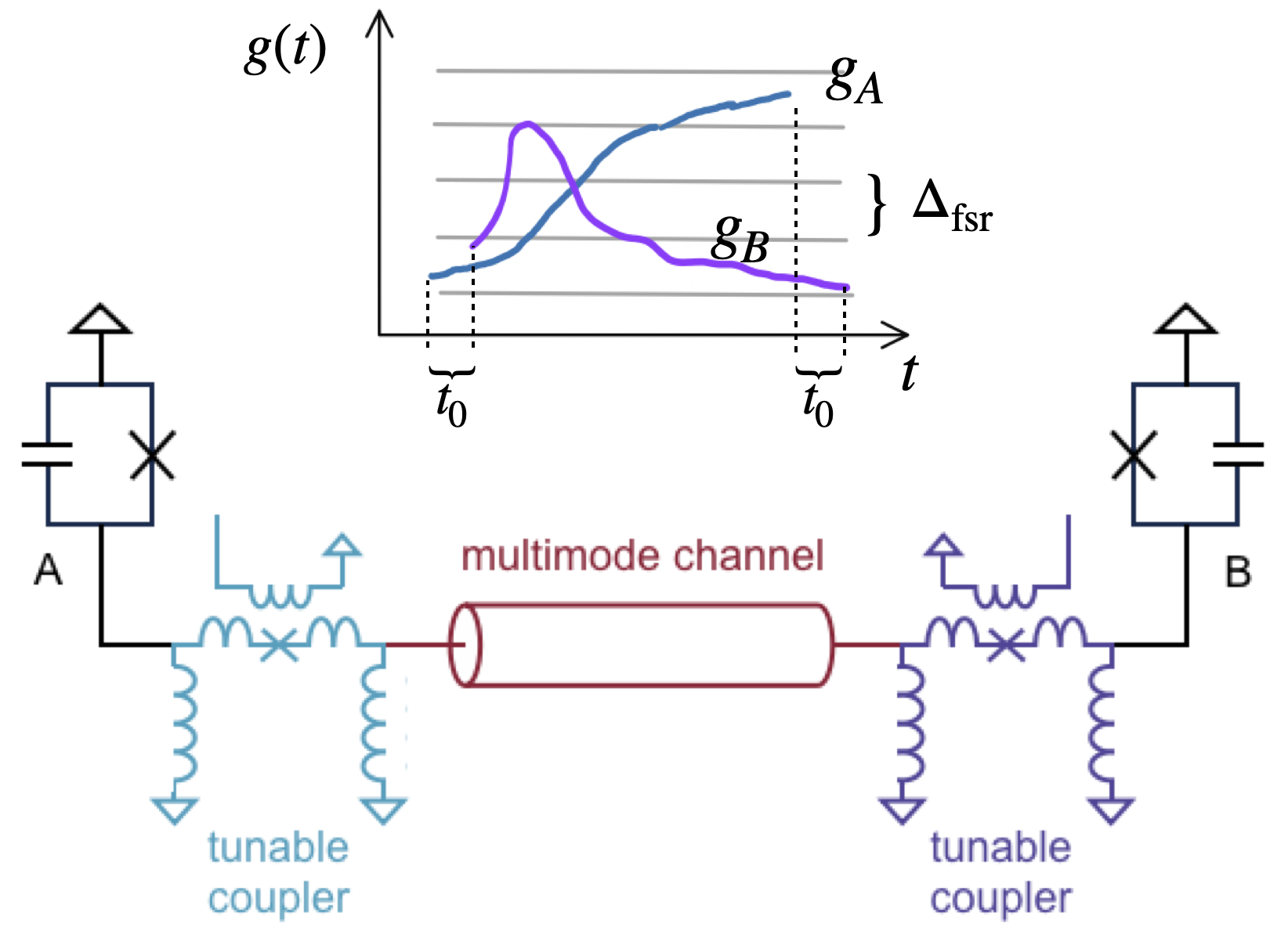} 
    \caption{Schematic of the settings. Two qubits (labeled by A and B)
    are connected by a channel through two tunable couplers. 
    The coupling $g_A(t)$ and $g_B(t)$ are activated sequentially with a delay of $t_0=\pi/\fsr$. Our scheme enables couplings with strength comparable with $\fsr$.}\label{figsys}
    \end{figure}

As depicted in Fig.{~}\ref{figsys}, the simplest setup of QST consists of two transmon qubits~\cite{Koch:2007aa} 
connected by a quantum channel. The channel supports standing waves
with free spectrum range $\fsr$, which is reversely proportional to
the channel length. To accurately model the system,
one must compare the qubit-channel coupling strength $g$ with $\fsr$. 
For \emph{short-range} channels where 
$\fsr\gg g$, only the mode closest to resonance is
relevant. The channel is thus modeled by a single-mode cavity and QST with high fidelity is achieved by 
methods based on Stimulated-Raman-Adiabatic-Passage (STIRAP)~\cite{Vitanov:2017aa,Bergmann:2019aa}.
For \emph{long-range} channels where $\fsr\rightarrow 0$, the channel modes can be 
viewed as a continuum. In this case, 
Cirac, Zoller, Kimble and Mabuchi (CZKM)~\cite{Cirac:1997aa} proposed a QST
protocol based on the input-output formalism and cascaded dynamics~\cite{Gardiner:1993aa,Carmichael:1993aa}.
It is also referred to as the \emph{pitch-and-catch} or
\emph{relay} method~\cite{Zhong:2019aa}.  The CZKM scheme has three steps
(1) pitch: qubit A emits a shaped photon; 
(2) the photon flies to qubit B;
(3) catch: B absorbs this photon through controlled interaction.

However, medium-range QST has not received sufficient theoretical attention
despite emerging experiments~\cite{McKay:2015aa,Sundaresan:2015aa,Chakram:2022aa,Chang:2020aa,Bienfait:2019aa,Zhong:2021aa,Niu:2023aa,Grebel:2024aa}. 
Transmon-waveguide couplings typically have
$g\lesssim 2\pi{ \times} 10^2\mathrm{MHz}$~\cite{Mirhosseini:2019aa,Sheremet:2023aa}.
A quick estimation shows that 
medium-range $\fsr$ is comparable to the attainable $g$ so that
the multimode nature is essential. In this regime, the old practice of the CZKM
scheme is questionable because the prerequisites of 
cascaded system and Markovian input-output formalism are inapplicable.
The STIRAP-based methods are also suboptimal 
because one has to suppress
$g$ intentionally to prevent the involvement of multiple channel modes
with a price of long scheme time hence the risks of qubit dissipation.
Such tradeoff stimulated ``hybrid'' method
employing STIRAP bright states to
trade channel-loss-immunity for
faster QST~\cite{Wang:2012aa,Bienfait:2019aa,Zhong:2021aa,Niu:2023aa}.
Nevertheless, to operate QST in the regime of $g\sim\fsr$,
we need to treat multiple channel modes coherently in a theory 
which goes beyond STIRAP and CZKM. 

In this Letter, we develop a new QST formalism  which
reduces to STIRAP and the CZKM scheme at the short and long-range limits, respectively, 
consolidating their benefits of 
low loss and high speed from each of them. 
We notice that dynamics in the regime of 
$g\,{\sim}\,\fsr$ is also interested in other fields~\cite{Krimer:2014aa,Han:2016aa,Johnson:2019aa,Lechner:2023aa,Lentrodt:2023aa}.

\paragraph*{System and Model.} We study the settings outlined in Fig.~\ref{figsys},
where two qubits (labeled by $A$ and $B$) are interconnected via a
channel using two tunable couplers~\cite{Chen:2014aa}.
The qubits have the same transition frequency $\omega_0$, and
the channel contains a set of equally-spaced (by $\fsr$) 
discrete levels with frequency $\{\omega_k\}_k$. The Hamiltonian
under the rotating-wave approximation reads
\begin{equation}\label{H}
    \begin{aligned}
    H(t)=  & g_A(t)\hat{\sigma}_A^\dagger  \sum_k  \hat{c}_k e^{-i\delta_k t} \\
    &\quad +g_B(t)\hat{\sigma}_B^\dagger\sum_k (-1)^k\hat{c}_k e^{-i\delta_k t}+ \text{H.c.}
    \end{aligned}
\end{equation}
where $\hat{\sigma}_{A/B}^\dagger$ is the qubit creation operator, 
$c_k$ denotes the channel annihilation operator,
$\delta_k=\omega_k-\omega_0$,
$\text{H.c.}$ represents Hermitian conjugate,
and the phase $(-1)^k$ in the second line reflects the distinct spatial 
profiles of even and odd channel
modes~\cite{Chang:2020aa,Malekakhlagh:2024aa}. QST is described by
the evolution of the state
\begin{equation}\label{state}
    \ket{\Psi(t)}=\alpha(t)\ket{1}_A+\beta(t)\ket{1}_B+\sum_k c_k (t)\ket{1}_k,
\end{equation}
where $\alpha(t)$, $\beta(t)$ and $c_k(t)$ are superposition coefficients and 
$\ket{1}_{A}$, $\ket{1}_{B}$ and $\ket{1}_{k}$ denote the state with 
a single excitation in qubit $A$,
$B$, and mode $k$, respectively. The initial state has $\alpha(t_i)=1$
and $\beta(t_i)=c_k(t_i)=0$. 
The target is to determine the profiles of $g_{A/B}(t)$ to 
achieve $\beta=1$ in the end.

To start, it is observed that A couples
to a collective operator $\sum_{k}\hat{c}_k e^{-i\delta_k t}$ in $H(t)$, and the same operator couples to B in $H(t+t_0$) 
with $t_0=\pi/\fsr$. This observation inspires us to delay 
$g_B(t)$ relative to $g_A(t)$ by $t_0$, cf. Fig.~\ref{figsys}, 
to ensure proper response of B to A. 
This retardation is further justified by causality:
For standard waveguides where
$\omega_k=k\fsr\,(\forall k\geq 1)$,
$t_0$ is exactly the time a photon takes to propagate
from A to B. We introduce a tilde over notations
to indicate the shift of the reference of time, e.g., $\tbeta(t)\equiv \beta(t+t_0)$.
It is convenient to apply the convention
that $g_A(t)=\tgb(t)=0$ for $t<t_i$
and $t>t_f$, where $t_f$ marks the completion of the control.
This convention leads to a formula for $c_k(t_f+t_0)$, or equivalently 
$\tilde{c}_k(t_f)$~\cite{sp},
\begin{equation}\label{ck}
    \tilde{c}_k(t_f) =-i\int_{t_i}^{t_f} d\tau \bigg[
        g^{*}_A(\tau)\alpha(\tau)+\tgb^{*}(\tau)\tbeta(\tau)\bigg]
    e^{i\delta_k \tau},
\end{equation}
where we have absorbed a factor of $e^{i\omega_0 t_0}$ into the definition 
of $\tgb$. QST is successful only if 
$\tilde{c}_k(t_f)=0,\,\forall k$. Note that Eq.~\eqref{ck}
vanishes as long as terms in the square bracket are orthogonal to 
$\exp{(i\delta_k\tau)}\,\forall k$ in the space of 
square-integrable functions over $\tau\in[t_i, t_f]$. Though not unique, 
the most appealing choice for the terms in the square bracket of Eq.~\eqref{ck}
might be simply zero,
\begin{equation}\label{adsc}
    g^{*}_A(\tau)\alpha(\tau)+\tgb^{*}(\tau)\tbeta(\tau)=0.
\end{equation}
Equation~\eqref{adsc} is different from the standard STIRAP dark state condition
by a delay of $t_0$ between the two terms.
Thus, we refer to it as the
\emph{asynchronous dark state condition} (ADSC).
A few lines of calculation shows that ADSC implies the channel mode formula~\cite{sp}
\begin{equation}\label{ckt}
    c_k(t)=-i\int_{t-t_0}^{t}d\tau g_A^{*}(\tau)\alpha(\tau) e^{i\delta_k \tau},
\end{equation}
and the equations for qubit amplitudes 
\begin{subequations}\label{eqab}
    \begin{equation}\label{eqa}
        \dot{\alpha}(t)  =-g_A(t)\int_{t-t_0}^{t}d\tau K(t-\tau) g_A^{*}(\tau)\alpha(\tau), 
    \end{equation}
    \begin{equation}\label{eqb}
        \dot{\tbeta}(t) =\tgb(t)\int_{t}^{t+t_0}d\tau K(t-\tau) \tgb^{*}(\tau)\tbeta(\tau),
    \end{equation}
\end{subequations}
where $K(t-\tau)=\sum_{k} e^{-i\delta_k (t-\tau)}$. 

These formulae display two notable features. Firstly, Eqs.~\eqref{eqa} and~\eqref{eqb} 
are formally decoupled: The interplay between the two qubits
disappears from the equations of motion.
This is unusual  because when the system is not cascaded
there are often complicated 
feed-back or forward written in the equations of motions, see
for example Refs.~\cite{Sinha:2020aa,Zhang:2023aa}. 
Secondly, we learn from the integral limits of Eqs.~\eqref{ckt} and~\eqref{eqab} 
that ADSC leads to a
finite memory time bounded by $t_0$, irrespective of the system specifications. 
We shall see below that these two pivotal features are the driving force behind every
aspect of our scheme.

\paragraph*{Interpolation between STIRAP and CZKM.} Here
we demonstrate that STIRAP and the CZKM scheme can be 
reproduced from our formalism. For STIRAP,
taking the short-channel limit $t_0\rightarrow 0$ formally reduces Eq.~\eqref{adsc} 
into the standard STIRAP dark state condition. However, the
conceptional difference is that, 
actually, $t_0=0$ is 
the limit of $\fsr\rightarrow\infty$ where all
channel modes are far blue detuned and no effective coupling occurs.
In contrast, we shall see that by keeping a finite $t_0$ explicitly, ADSC~\eqref{adsc} can 
be satisfied without the adiabatic approximation,
which underlies the standard
STIRAP and results in slow evolutions~\cite{Baksic:2016aa,Malekakhlagh:2024aa}.

The CZKM scheme uses cascaded setups 
and Markovian channels where $K(t)\propto \delta(t)$. Here,
the formal decoupling of Eqs.~\eqref{eqa} and~\eqref{eqb} is akin to 
cascaded dynamics. Beyond the Markovian limit,
our ADSC-based formalism extends the validity of the CZKM pulses to
a regime having not been envisioned. That is, if there are many discrete modes 
below and above the qubit frequency $\omega_0$ and the coupling is not too strong,
the kernel $K(t)$ would be approximated by~\cite{Milonni:1983aa}
\begin{equation}\label{delta}
    K(t)\propto \lim_{N\rightarrow\infty}\sum_{k=-N}^{N}e^{-ik\fsr t}
    \propto 2t_0 \sum_{n=-\infty }^{\infty} \delta(t-2n t_0),
\end{equation}
Note that only the term of $n=0$ in the summation survives when the above formula
is substituted into Eqs.~\eqref{eqab}. In this case, Eq.~\eqref{eqab} is 
effectively Markovian. It explains the numerical observation that 
the CZKM scheme works well in some intermediate regimes~\cite{Penas:2022aa}.

\paragraph*{General Solutions.} Now we are in a position to find
the solutions of Eq.~\eqref{eqab} satisfying ADSC. 
As functions of time, $g_A(t)$ and $g_B(t)$ have infinitely many degrees of freedom. 
ADSC is a constraint that reduces these degrees of freedom by half.
Thus, one strategy is to choose an arbitrary $g_A(t)$
and then find the corresponding $g_B(t)$. The key factor for this
strategy to work is the formal decoupling of Eq.~\eqref{eqa} and Eq.~\eqref{eqb}.
To proceed, for any given $g_A(t)$, we substitute it into Eq.~\eqref{eqa}
and obtain the trajectory $\alpha(t)$. This 
$g_A(t)$ is accepted as long as 
$\alpha(t_f)=0$ (up to an exponentially small correction).
Suppose that such a pair of $[g_A(t), \alpha(t)]$ is obtained. The 
product $x(t)\equiv g_A^*(t)\alpha(t)$
is thus known and ADSC implies that $\tgb^*(t)\tbeta(t)=-x(t)$.  Next, we define
the abbreviation $y(t)\equiv -\dot{\tbeta}(t)/\tgb(t)$ which is determined by
\begin{equation}
    y(t)=\int_{t}^{t+t_0}d\tau\, K(t-\tau)x(\tau).
\end{equation}
Then Eq.~\eqref{eqb} can be rephrased as 
\begin{equation}\label{eqb-2}
    \tbeta^{*}(t)\dot{\tbeta}(t)=x^{*}(t)y(t),
\end{equation}
The real part of Eq.~\eqref{eqb-2} implies that
\begin{equation}\label{absbeta}
    \frac{d}{dt}\abs{\tbeta(t)}^2=2\,\mathrm{Re}\big[x^{*}(t)y(t)\big],
\end{equation}
which completely determines the magnitude of
$\tbeta(t)$. 
The imaginary part of Eq.~\eqref{eqb-2} implies that
\begin{equation}
    \abs{\tbeta(t)}^2\frac{d}{dt}\arg \tbeta(t)=\mathrm{Im}\big[x^*(t)y(t)\big],
\end{equation}
which fixes the phase of $\tbeta(t)$. Now $\tbeta(t)$ is 
known and $\tgb(t)$ is simply $-x^*(t)/\tbeta^*(t)$. One can
verify the ADSC results by substituting the pair of $[g_A(t),\tgb(t)]$ into 
the original Schr\"{o}dinger equation.

In the end, the problem is reduced to the general existence of such
$g_A(t)$ so that Eq.~\eqref{eqa} produces $\alpha(t_f)\approx 0$. 
We cannot prove it rigorously but merely argue the existence using the constant ansatz 
$g_A(t)=g$. [To satisfy ADSC, $g_A(t)$ must ramp up from zero.
If the ramping is quick, $g_A(t)$ can be approximately viewed as 
a constant.]
We assume $t_i=0$ for brevity and obtain the inverse Laplace transformation
of Eq.~\eqref{eqa}
\begin{equation}\label{laplace}
\alpha(t)=\int_{\epsilon-i\infty}^{\epsilon+i\infty}\frac{e^{st} ds}{2\pi i}
 \bigg[s+g^2\sum_k\frac{1-e^{-(s+i\delta_k)t_0}}{s+i\delta_k}\bigg]^{-1},
\end{equation}
where $\epsilon>0$. We find that, 
due to the finite memory time $t_0$ of Eq.~\eqref{eqa}, all
the poles of the integrand have negative real parts~\cite{sp}. 
Thus, according to Cauchy's residue theorem, $\alpha(t)$
is a sum of exponential decays that vanishes eventually. We
have verified that such exponential decay can be fast enough~\cite{sp}.

To compare, if only qubit A is coupled to the channel,
following the step of ``pitch'' 
of the CZKM scheme, the counterpart of Eq.~\eqref{eqa} 
will include the full history of $\alpha(t)$.
Now all the poles of its Laplace transformation locate on the 
imaginary axis~\cite{sp} so that $\alpha(t)$ 
is a sum of harmonic oscillations.

\paragraph*{Robustness to noises.} 
QST is affected by the losses from both the qubits and the channel.
Denote the qubit loss rate  
by $\gamma$ and the channel loss rate by $\kappa$.
Generally we have $\gamma < \kappa$ and
the infidelity of QST is estimated by
$\gamma T+(\kappa-\gamma)\mathcal{E}$,
where $T=t_f-t_i+t_0$ is the scheme time, and $\mathcal{E}$ 
is integrated channel population~\cite{sp}
\begin{equation}\label{loss}
    \begin{aligned}
   \mathcal{E}\equiv & \sum_k\int_{t_i}^{t_f+t_0}\,\abs{c_k(t)}^2 dt \\
        = &  t_0 + \int_{t_i}^{t_f} \tau 
        \frac{d}{d\tau}\big[ \abs{\alpha(\tau)}^2+\abs{\tbeta(\tau)}^2 \big] d\tau.
    \end{aligned} 
\end{equation}
To protect QST from the losses, smaller $T$ and $\mathcal{E}$ are favored.
It is conceivable that $T$ can be much smaller than the 
STIRAP-based methods because our scheme allows stronger couplings $g\sim\fsr$.
As to $\mathcal{E}$, we find it remains small
for any realizations of $g_A(t)$ and $g_B(t)$ 
as long as ADSC is obeyed~\cite{sp}. Briefly, by substituting
Eqs.~\eqref{adsc} and~\eqref{eqab} into Eq.~\eqref{loss}, we 
find that 
the second term of Eq.~\eqref{loss} scales as
$O(t_{*})$, where $t_*$ is 
the memory time of Eq.~\eqref{eqab}, i.e., the smaller one of
$t_0$ (imposed by ADSC)
and the intrinsic memory time determined by the channel kernel $K(t)$. 
For medium-range channels, we have $t_*=t_0$ thus 
$\mathcal{E}=O(t_0)$. It is important that here $T$ is not relevant to $\mathcal{E}$
so that the tradeoff between $T$ and $\mathcal{E}$~\cite{Wang:2012aa} 
is resolved in our QST formalism.
For channels with Makovian kernel
($t_*=0$), Eq.~\eqref{loss} reproduces the loss of
flying photon $\mathcal{E}= t_0$. 

For realistic superconducting circuits, QST will be
affected by qubit level leakage~\cite{Sank:2016aa,Battistel:2021aa,Varbanov:2020aa,Lacroix:2023aa,McEwen:2021aa} and 
the dissipation from qubits and channels.
The robustness of our protocol against these issues will be demonstrated numerically in the Supplemental Material~\cite{sp}.
Particularly, our QST scheme is also immune to
thermal channel occupations if using harmonic 
oscillators as the sender and receiver, same with Refs.~\cite{Vermersch:2017aa,Xiang:2017aa}
but in a broader regime.

\begin{figure}[b]
    \centering
    \includegraphics[width=\textwidth]{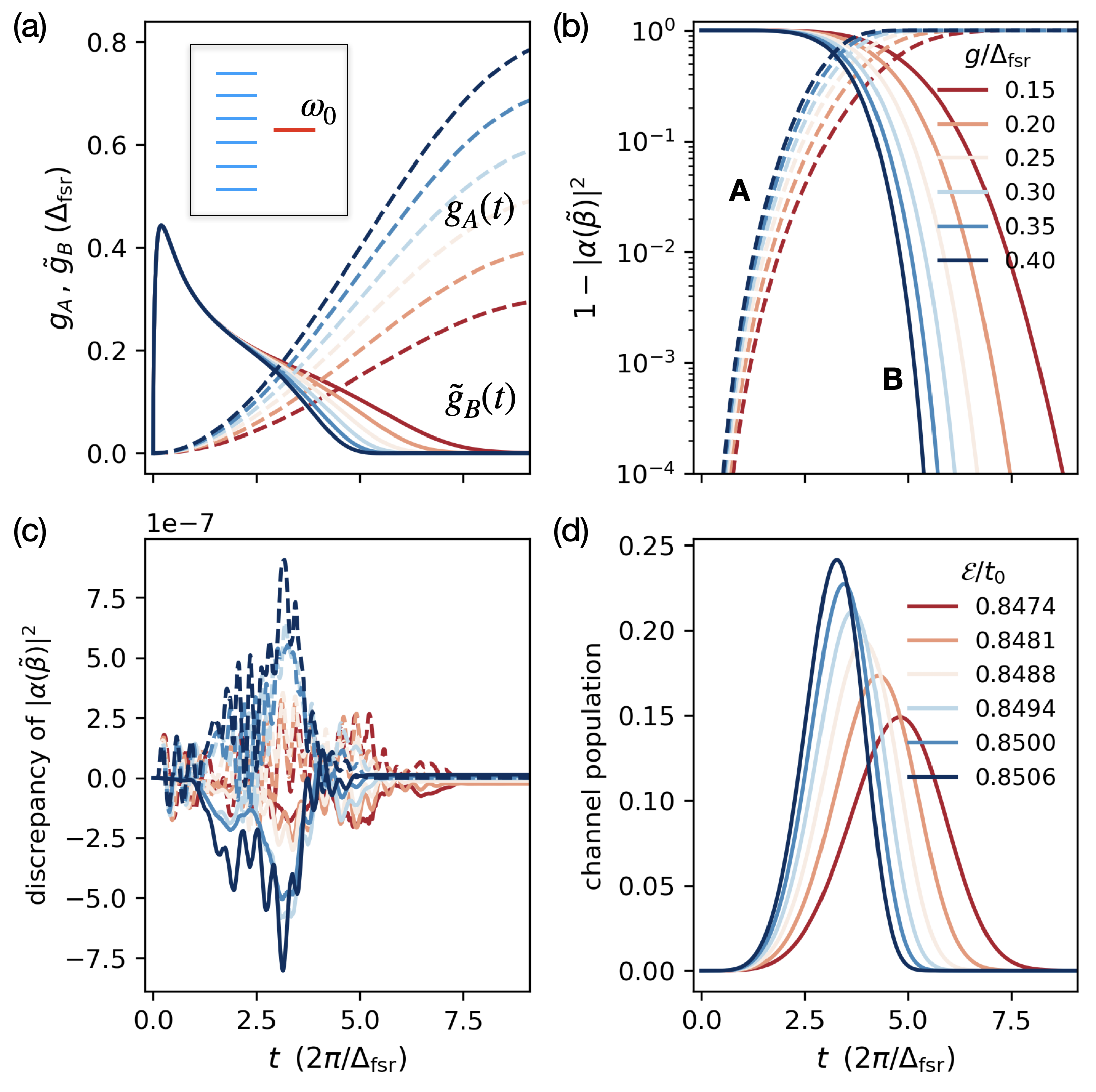} 
    \caption{ Numerical results for the setup with level structure depicted in the inset of (a),
    which contains six channel modes.  
    (a) $g_A$ (dashed curves) and the companion $\tgb$ (solid). 
    (b) Reversed populations $1-\abs{\alpha}^2$ (dashed curves) and $1-|\tbeta|^2$ (solid curves) 
    calculated from the original Schr\"{o}dinger equation.
    (c) The discrepancies between the original Schr\"{o}dinger equation
    and the ADSC-based equations: Dashed curves for $\abs{\alpha}^2$, and 
    solid curves for $|\tbeta|^2$.
    (d) Channel populations and the
    table of the integrated population $\mathcal{E}/t_0$. 
    }
    \label{fig2}
    \end{figure}

\begin{figure}[t]
    \centering
    \includegraphics[width=\textwidth]{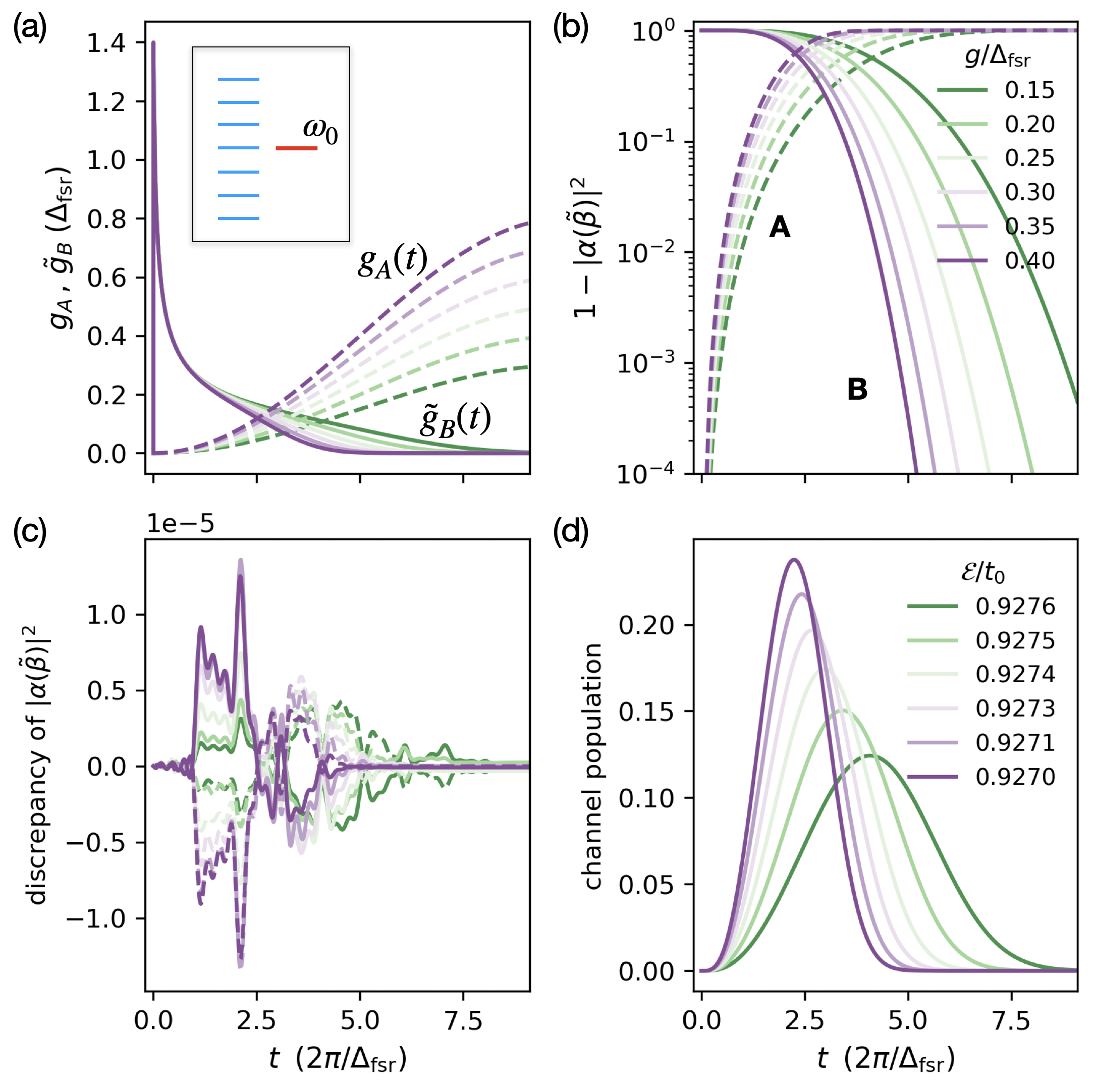} 
    \caption{Numerical results for the setup with level structures depicted in the inset of 
    (a), which contains seven channel modes. The subplots (a-d) are arranged in
    the same manner of Fig.~\ref{fig2}.}
    \label{fig3}
    \end{figure}

\paragraph*{Examples.} As a demonstration of principle, 
we accommodate ourselves by considering two scenarios where 
the kernel $K(t)$, control pulses $g_{A/B}(t)$ and 
amplitudes $\alpha(\beta)$ are all real-valued. In Case 1, $\omega_0$ is 
in the middle of two channel modes (detuned by $\pm\fsr/2$), see 
the inset of Fig.~\ref{fig2}(a). In Case 2,
$\omega_0$ is resonant with one channel mode,
see the inset of Fig.~\ref{fig3}(a). Each case
involves an equal number of channel modes above 
and below $\omega_0$, set here to be three. We unintentionally
choose $g_A(t)=g+g\sin[(t-5)/(10\pi)]$, where $g/\fsr=0.15\sim 0.4$ 
(in increments of 0.05, distinguished by colors) and $t$ denotes time 
in units of $2\pi/\fsr$ (thus $t_0=0.5$). Our computations utilize
the Python IDEsolver package~\cite{Karpel:2018aa} to solve 
Eq.~\eqref{eqa} and determine $\tgb(t)$. We then validate the control pulses 
by solving the original Schr\"{o}dinger equation using QuTiP~\cite{Johansson:2012aa,Johansson:2013aa}, and compare the 
results of $\alpha(t)$ and $\tbeta(t)$ with the predictions 
of ADSC-based formulae.
Numerical results for these two scenarios are 
shown in Figs.~\ref{fig2} and~\ref{fig3}, respectively. 

In Fig.~\ref{fig2}(a) we plot the shapes of $g_A(t)$ (dashed curves) 
and the companion $\tgb(t)$ (solid curves) obtained 
by the method introduced above. We then substitute the control pulses
into the original Schr\"{o}dinger equation and obtain 
the results of $1-|\alpha(t)|^2$ and $1-|\tbeta(t)|^2$ depicted in 
Fig.~\ref{fig2}(b). The curves confirm that stronger $g_A$ facilitates faster 
QST. The results shown in
Fig.~\ref{fig2}(b) are compared with results calculated by Eqs.~\eqref{eqa}
and~\eqref{absbeta}. Their 
discrepancies are found to be smaller than $10^{-6}$, as
plotted in Fig.~\ref{fig2}(c). This error be further reduced 
by decreasing the temporal
step length (set at $\Delta t=0.002$) in the numerical calculations.
To assess the risk of channel loss, 
we plot the channel population $\sum_k \abs{c_k(t)}^2$ in Fig.~\ref{fig2}(d),
accompanied by the integrated population $\mathcal{E}$~\eqref{loss}
for each $g/\fsr$. The results corroborate our analysis
that $\mathcal{E}= O(t_0)$, and, somewhat surprisingly, 
$\mathcal{E}$ is almost invariant with respect to $g/\fsr$. 

Figure~\ref{fig3} conveys the similar message. 
Notably, the control sequences  
$\tgb(t)$ shown in Fig.~\ref{fig3}(a) exhibit higher peaks. 
The discrepancies between the full Sch\"{o}dinger equation and ADSC predictions 
are at the level of $10^{-5}$ (obtained with $\Delta t=0.001$), 
cf. Fig.~\ref{fig3}(c). The integrated channel population is
$0.927$ for all values of $g/\fsr$, exceeding than 
that observed in Fig.~\ref{fig2}(d) by approximately 10\%. This may 
elucidate the estimation that 
the level configuration of Case 2 yields higher fidelity~\cite{Teoh:2023aa}.

\paragraph*{Conclusion.} In this Letter, we have developed a formalism 
for QST through a multimode channel. This central idea is to control 
the system time-dependently to satisfy ADSC~\eqref{adsc}. Our approach
integrates key features of both STIRAP and the CZKM scheme, 
enabling QST implementation with coupling strengths comparable to the 
free spectral range of the channel $g\sim\fsr$. 
The increased coupling strengths give rise to 
faster QST, which is favored for minimizing qubit loss. 
Meanwhile, the risk of channel loss scales as
$O(\kappa t_0)$, where $t_0=\pi/\fsr$, independent of system specifications. 
The level of loss is acceptable because $t_0$ is typically much smaller
than the QST scheme time. Our results have potential applications in
the realization of fast and reliable chip-to-chip quantum communication,
entanglement distribution,
and couplers between non-neighboring qubits, which are crucial for large 
scale fault-tolerant quantum computation. For future works,
one may propose other strategies for solving Eq.~\eqref{eqab} 
and optimize the control pulses under realistic conditions.
Remote quantum logic gates and frequency/time division 
multiplexing based on ADSC~\eqref{adsc} might also be interesting topics.

\begin{acknowledgements}
    Y.-X. Z. thanks the discussions with Y. Zhong and the comments from J. J. Garc\'{i}a-Rippoll,
    and acknowledges the financial support from
    Innovation Program for Quantum Science and Technology (Grant No. 2023ZD0301100 and No.~2-6),
    National Natural Science Foundation of China (Grant No.~12375024), and
    CAS Project for Young Scientists in Basic Research (YSBR-100).
\end{acknowledgements}

\bibliography{StateTransfer.bib}

\end{document}

\title{ Supplemental Material to ``Quantum State Transfer via a Multimode Resonator''}
\author{Yang He}
\affiliation{Institute of Physics, Chinese Academy of Sciences, Beijing 100190, China}
\affiliation{School of Physical Sciences, University of Chinese Academy of Sciences, Beijing 100049, China}
\author{Yu-Xiang Zhang}
\affiliation{Institute of Physics, Chinese Academy of Sciences, Beijing 100190, China}
\affiliation{School of Physical Sciences, University of Chinese Academy of Sciences, Beijing 100049, China}
\affiliation{Hefei National Laboratory, Hefei, 230088, China}
\date{\today}

\begin{abstract}
The Supplemental Material is arranged as follows. In Sec.~\ref{Sec:ADSC} we derive Eqs.~(5,6) of the
main text. In Sec.~\ref{Sec:Laplace} we discuss the Laplace transformation 
of the integro-differential equation~(6a) of the main text given a constant $g_A(t)=g$. 
In Sec.~\ref{Sec:Loss} we present the details for the claim of $\mathcal{E}=O(t_0)$.
In Sec.~\ref{Sec:Simulate} we numerically study the adverse effects of qubit leakage error
and qubit/channel losses. In Sec.~\ref{Sec:Thermal} we study the robustness to thermal channel
photon population. 
\end{abstract}

\maketitle

\section{Schr\"{o}dinger Equations and ADSC}\label{Sec:ADSC}
Here we derive the equations of motion using the Hamiltonian and 
single-excitation ansatz introduced in the main text.
Schr\"{o}dinger equation for the channel mode amplitude $c_k$ reads 
\begin{equation}
    \dot{c}_k(t)=-i \alpha(t) g^{*}_A(t)e^{i\delta_k t}
    -i\beta(t)(-1)^k g^{*}_B(t)e^{i\delta_k t}.
\end{equation}
Using the initial condition $c_k(t_i)=0$, 
the above equation is formally solved by
\begin{equation}\label{S-ck1}
    c_k(t)  =-i \int_{t_i}^t d\tau \big[ g^{*}_A(\tau)\alpha(\tau)
    +(-1)^k  g^{*}_B(\tau){\beta}(\tau) \big]e^{i\delta_k \tau}.
\end{equation}
Now let us replace $ g^{*}_B(\tau){\beta}(\tau) $ by 
$ \tgb^{*}(\tau-t_0)\tbeta(\tau-t_0)$ so that the second part is rewritten as 
\begin{equation}
(-1)^k\int_{t_i-t_0}^{t-t_0} d\tau \tgb^{*}(\tau){\tbeta}(\tau) e^{i\delta_k(\tau+t_0)}
\end{equation}
Recall the convention that $\tgb(t)=0$ for $t< t_i$ and 
\begin{equation}
    e^{i\delta_k t_0}= (-1)^k e^{-i\omega_0 t_0}.
\end{equation}
We can also choose $e^{i\delta_k t_0}= (-1)^{k+1} e^{-i\omega_0 t_0}$ on the right hand side of 
the above formula, because $k$ is just a numbering of the channel modes.
Equation~\eqref{S-ck1} is then 
\begin{equation}\label{S-ck2}
    \begin{aligned}
    c_k(t)=& -i \int_{t_i}^{t}d\tau g^{*}_A(\tau)\alpha(\tau)e^{i\delta_k \tau} \\
    &\qquad -i e^{-i\omega_0 t_0}\int_{t_i}^{t-t_0} d\tau \tgb^{*}(\tau){\tbeta}(\tau)e^{i\delta_k \tau}.
    \end{aligned}
\end{equation}
To obtain Eq.~(3) 
of the main text, one just needs to substitute $t=t_f+t_0$ into the 
above formula and use the convention that $g_A(t)=0$ for $t> t_f$,
and redefine $\tgb(t)$ by $e^{i\omega_0 t_0}\tgb(t)$.

Equations for the amplitudes of the two qubits are
\begin{subequations}
    \begin{equation}
        \dot{\alpha}(t)  =-ig_A(t)\sum_k e^{-i\delta_k t}c_k(t) 
    \end{equation}
    \begin{equation}\label{S-eqb1}
        \dot{\beta}(t) =-ig_B(t)\sum_k (-1)^k e^{-i\delta_k t}c_k(t)
    \end{equation}
\end{subequations}
For qubit A, we substitute Eq.~\eqref{S-ck2} into the above formula and obtain 
\begin{equation}
        \begin{aligned}
        \dot{\alpha}(t) = & -g_A(t)\int_{t_i}^t d\tau K(t-\tau) g^{*}_A(\tau)\alpha(\tau) \\
         &- g_A(t)\int_{t_i}^{t-t_0}d\tau 
        K(t-\tau)\tilde{g}^{*}_B(\tau)\tilde{\beta}(\tau) e^{-i\omega_0 t_0}
        \end{aligned}
    \end{equation}
Then ADSC cancels the integral over $[t_i, t-t_0]$ hence results in 
Eq.~(6a) of the main text. The case of qubit B is similar.

\section{Distribution of the Poles}\label{Sec:Laplace}
The poles of Eq.~(12) of the main text are determined by solving
the following equation 
\begin{equation}
    s+g^2\sum_k\frac{1-e^{-(s+i\delta_k)t_0}}{s+i\delta_k}=0.
\end{equation}
We change the variable as $s\rightarrow i(z+\omega_0)$ and obtain
\begin{equation}
    z+\omega_0-g^2 \sum_k \frac{1-(-1)^k e^{iz t_0}}{z+ k\fsr}=0.
\end{equation}
If there is a real solution of $z$, the imaginary and real parts of the 
above formula both vanish,
\begin{subequations}
    \begin{equation}\label{S-real}
        \sin(zt_0)\sum_k \frac{(-1)^k}{z+k\fsr}=0
    \end{equation}
    \begin{equation}\label{S-imag}
        z+\omega_0 -g^2\sum_k \frac{1-(-1)^k \cos(zt_0)}{z+k\fsr}=0.
    \end{equation}
\end{subequations}
Equation~\eqref{S-real} implies that either $\sin(zt_0)=0$ or the summation vanishes.
Nevertheless, it vanishes only for a countable set of values of $z$,
which are independent to $\omega_0$. However, $\omega_0$ appears in Eq.~\eqref{S-imag}
so that Eqs.~\eqref{S-real} and~\eqref{S-imag} are both fulfilled 
only if $\omega_0$ equals 
some certain values. Thus, in general 
there are no real solutions to $z$, equivalently, no purely 
imaginary poles of the Laplace transformation, which means that QST will succeed
up to an exponentially decreasing error.

\begin{figure}[tbh]
    \centering
    \includegraphics[width=0.85\textwidth]{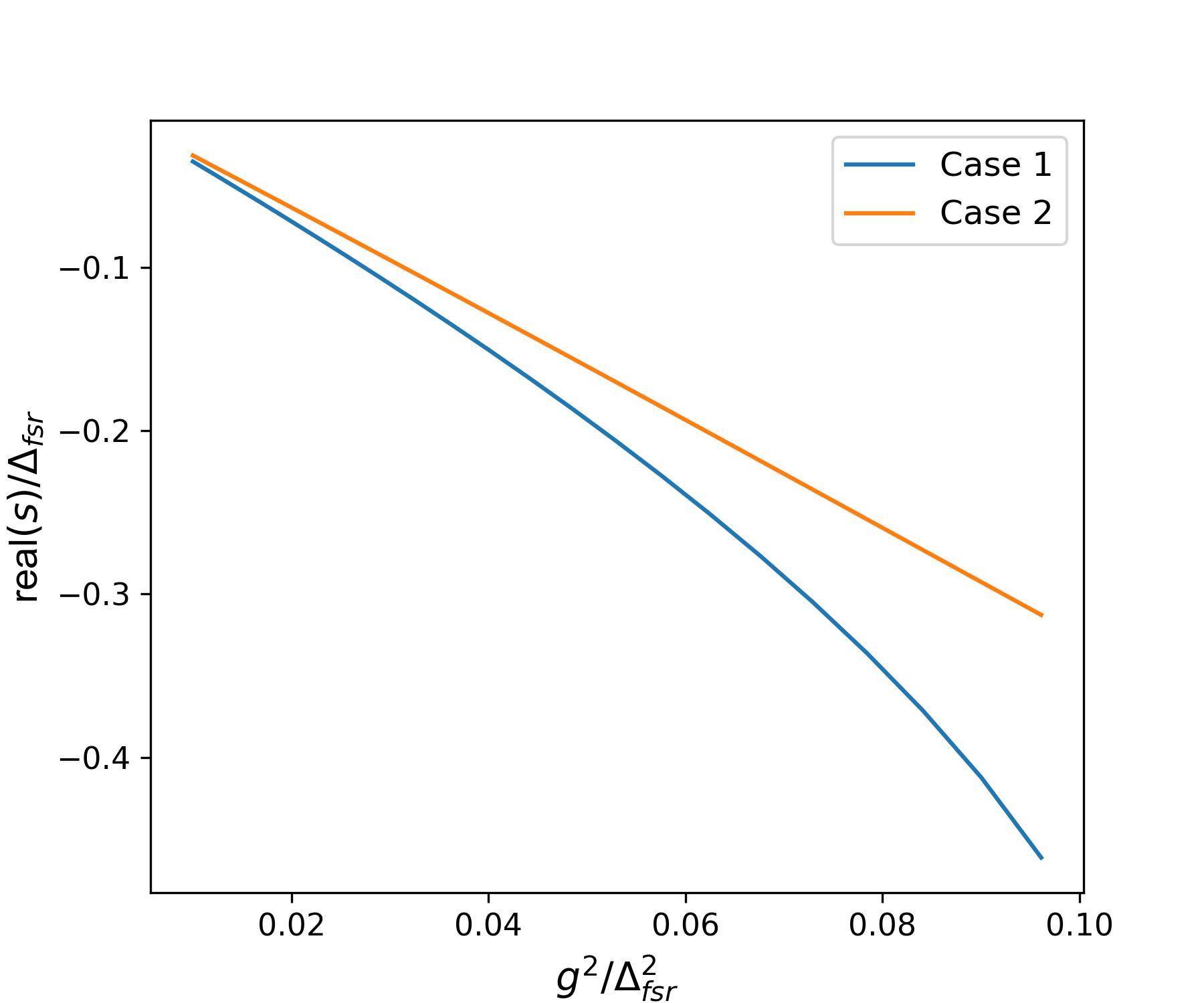} 
    \caption{The largest real part of the poles (in the unit of $\fsr$) of the 
    Laplace transformation formula of the two cases introduced in the
    main text.} \label{sp_reals}
\end{figure}

To have an idea of how fast the exponential decay could be, we plot the slowest
decay rate (largest real part of the poles $s$, which are negative) 
for the two examples studied in the main text in Fig.~\ref{sp_reals}. 
The plot shows that QST can be efficiently implemented provided that the
coupling strength $g$ is not too small.

The above result can be compared with the integro-differential equation
where the full history is included, which is exactly the equation describing the interaction between a single qubit
and the channel reads 
\begin{equation}
    \dot{\alpha}(t)=-g^2\int_{0}^{t}d\tau K(t-\tau) \alpha(\tau).
\end{equation}
Provided that $\alpha(0)=1$, $\alpha(t)$ is solved by 
\begin{equation}\label{S-poles}
    \alpha(t)=\int_{\sigma-i\infty}^{\sigma+i\infty}\frac{e^{st} ds}{2\pi i}
 \bigg[s+g^2\sum_k\frac{1}{s+i\delta_k}\bigg]^{-1}.
\end{equation}
To find the poles, we change the variable as $s\rightarrow iz$ and obtain
\begin{equation}
    z-g^2\sum_k \frac{1}{z+\delta_k}=0.
\end{equation}
As an equation with real coefficients, if $z$ is a solution, so is $z^{*}$.
One of the corresponding $s$ of this conjugate pairs will have a positive real part.
This is impossible because the Laplace transformation of a bounded function 
is analytic on the right half of the complex plane.
Therefore, all the poles of Eq.~\eqref{S-poles} are distributed on the imaginary axis.

\section{Channel Loss}\label{Sec:Loss}
Here we derive the formula of channel loss. The normalization of 
state (2) of the main text simply implies that 
\begin{equation}
    \begin{aligned}
    \mathcal{E}= & \int_{t_i}^{t_f+t_0}dt\sum_k\abs{c_k(t)}^2 \\
    = & t_f-t_i+t_0-\int_{t_i}^{t_f+t_0}d\tau \big[
        \abs{\alpha(\tau)}^2+\abs{\beta(\tau)}^2
        \big]\\
    = & \int_{t_i}^{t_f+t_0} d\tau\;
    \tau\frac{d}{d\tau}\big[
        \abs{\alpha(\tau)}^2+\abs{\beta(\tau)}^2
        \big],
    \end{aligned}
\end{equation}
where we have used the formula of integration by parts in the third line.
Next, the conventions that $g_A(t>t_f)=0$ means that 
\begin{equation}
    \int_{t_i}^{t_f+t_0} d\tau\;
    \tau\frac{d}{d\tau}
        \abs{\alpha(\tau)}^2=\int_{t_i}^{t_f} d\tau\;
        \tau\frac{d}{d\tau}
            \abs{\alpha(\tau)}^2.
\end{equation}
Similarly, we can replace $\beta(t)$ by $\tbeta(t)$ and the convention that
$\tgb(t<t_i)=0$ implies that 
\begin{equation}
    \int_{t_i}^{t_f+t_0}d\tau\; \tau\frac{d}{d\tau} \abs{\beta(\tau)}^2
    = t_0+ \int_{t_i}^{t_f}d\tau\;\tau\frac{d}{d\tau} \abs{\tbeta(\tau)}^2.
\end{equation}
In the end, we obtain
\begin{equation}\label{S-P}
    \mathcal{E} =t_0+\int_{t_i}^{t_f}dt \; t\frac{d}{dt}\big[
        \abs{\alpha(t)}^2+\abs{\tbeta(t)}^2
        \big].
\end{equation}
We refer to the unraveled integral as ``(buffer)''  because
it vanishes if the emission from qubit A and the 
absorption of qubit B have the same pace.

Applying the Schr\"{o}dinger equation, ADSC, and 
the conventions that $g_A(t<t_i)=0$ and $\tgb(t>t_f)=0$, 
we can expand the (buffer) term into
\begin{equation}
    \begin{aligned}
 (\text{buffer}) 
    = &  \int_{-\infty}^{\infty}dt \int_{t-t_0}^{t}d\tau
    (t-\tau) \alpha^*(t)g_A(t) 
    \\ &\qquad \times K(t-\tau)  \alpha(\tau)g^{*}_A(\tau) + \text{c.c.},
    \end{aligned}
\end{equation}
where ``c.c.'' is a shorthand of complex conjugate.
Therein, let us assume 
When $t_{0}$ is much shorter than the scheme time $t_f-t_i$, 
we can view $t_0$ as a coarse grained 
scale of time within which
$g_A(t)$ and $\alpha(t)$ change little. Then, we simply replace the factor $(t-\tau)$
in the first line of the above formula with $t_0$. This is fine because
$0\leq t-\tau\leq t_0$ by definition. It leads to
\begin{equation}
    \begin{aligned}
        (\text{buffer})\sim 
 & t_0 \int_{-\infty}^{\infty}dt \int_{t-t_0}^{t}d\tau K(t-\tau)\\
 &\qquad \qquad \times \alpha^*(t)g_A(t) \alpha(\tau)g_A^*(\tau) + \text{c.c.} \\
 = & -t_0\int_{-\infty}^{\infty}dt \frac{d}{dt}\abs{\alpha (t)}^2\\
 =& t_0.
    \end{aligned}
\end{equation}
Therefore, the channel loss $\kappa P$, cf. Eq.~\eqref{S-P},
is bounded by the scale $O(\kappa t_0)$.

As a remark, in order to extend the above argument to long channels,
one should keep in mind that the  
effective memory time $t_{*}$ of the system is 
\begin{equation}
    t_{*}=\min\{ t_0, t_{\text{channel}} \},
\end{equation}
where $t_{\text{channel}}$ is the intrinsic memory time determined by 
the kernel function $K(t\gg t_{\text{channel}})\rightarrow 0$.
Generally speaking, we have $t_{*}=t_{\text{channel}}$ for long channels,
especially when the Markovian 
approximation works well. Thus, we should replace 
$t_0$ by $t_{*}$ in the estimation of the (buffer) term thereof.

\section{The Robustness to Qubit Leakage and Qubit/Channel Loss}\label{Sec:Simulate}

In this section, we study how the extra excitation in the 
sender initialization and the qubit and channel dissipation
affect the efficiency of our QST scheme. The figure of merit used to
characterize the influence is the inefficiency defined by
\be
1-\abs{\tbeta(t_f)}^2,
\ee
where that $\tbeta$ denotes the amplitude of having only a single excitation at B. 
Inefficiency defined above can also be understood as a measure of QST infidelity.

\subsection{Qubit Leakage to Higher Levels}
To account for the leakage to higher transmon levels,  we set the
transmon anharmonicity at 2.5 $\fsr$ and write the initial state 
of qubit A as
$$(1-\mathrm{er})^{0.5}\ket{1}_A+\mathrm{er}^{0.5}\ket{2}_A,$$ 
where  ``$\mathrm{er}$'' quantifies the leakage error
in sender's initial state preparation. Here a coherent leakage is assumed so that
qubit A is initialized in a pure state.
Channel and qubit B are assumed
to be in their ground states. We continue to use the controls $g_A(t)$ and $g_B(t)$
obtained for the ideal systems.

\begin{figure}[t]
    \centering
    \includegraphics[width=1\textwidth]{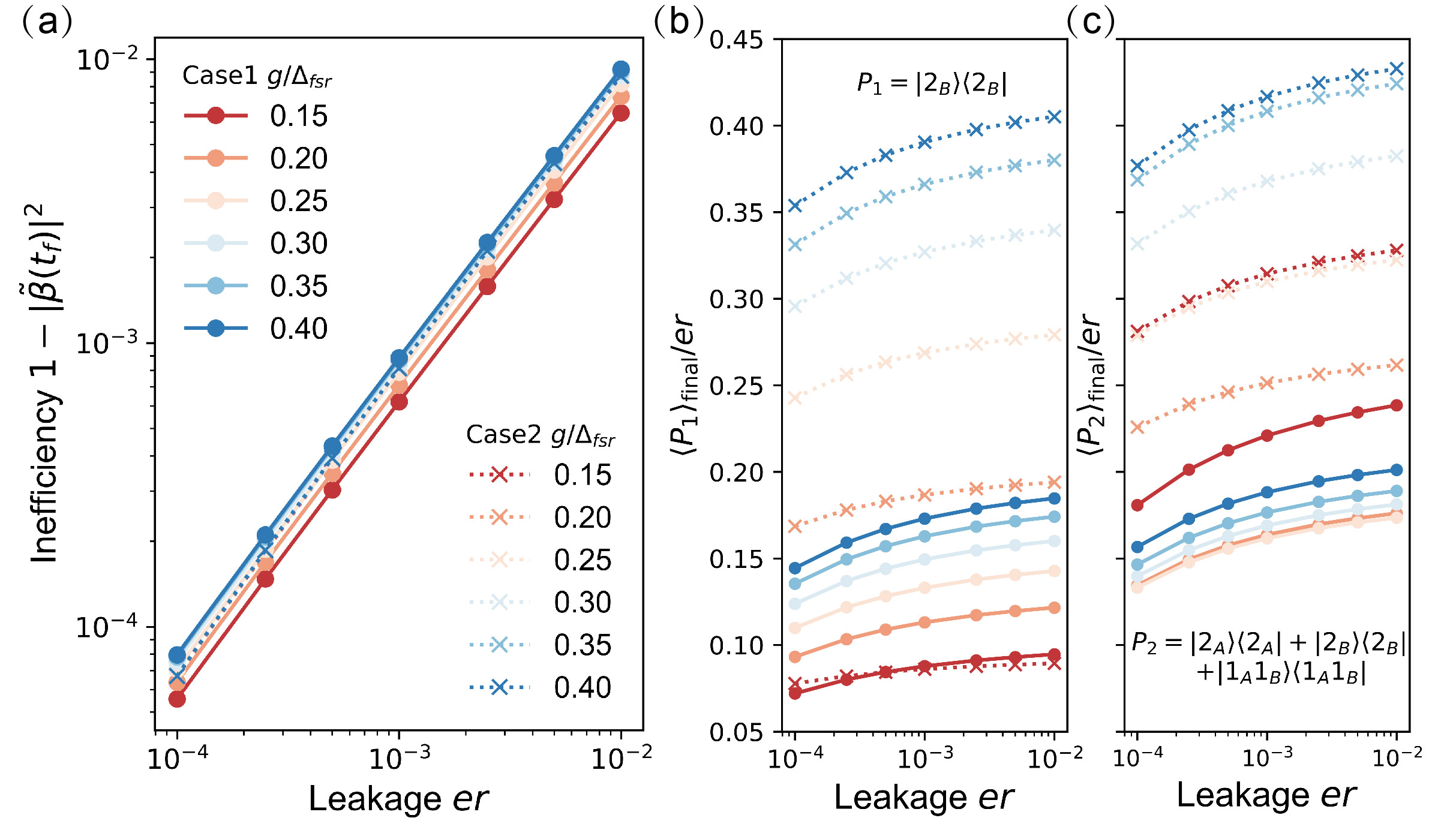} 
    \caption{QST inefficiency due to qubit leakage error. (a) Infidelity of QST, $1-\abs{\tbeta(t_f)}^2$,
    as the function of initial amount of leakage into $\ket{2_A}$, for the two examples introduced
    in the main text.  
    The solid curves represent Case 1 
    where the qubit frequency lies in the middle of two channel modes.
The dashed ones represent Case 2 where the qubits are in resonance 
    with one channel mode.
    (b) The population (relative to the initial leakage $\mathrm{er}$) of $\ket{2_B}$
    at the end of QST. (c) The final population (relative to the initial leakage  $\mathrm{er}$)
    on $\ket{2_A}$, $\ket{2_B}$ and $\ket{1_A}\otimes \ket{1_B}$.
    } \label{leak}
\end{figure}

We solve the system evolution using QuTiP and plot
QST inefficiency as a function of $\mathrm{er}$ in Fig.~\ref{leak}(a). It shows 
a linear relation between them, indicating the applicability of
first order perturbation. Since single qubit gates with an average 
gate error below $10^{-4}$ has been realized in Ref.~\cite{Li:2023aa},
we believe that qubit leakage should not be problematic in our scheme.

To see the location of the population leakage, we plot the final
population on $\ket{2_B}$ (normalized by the factor $\mathrm{er}$) in Fig.~\ref{leak}(b).
The results show that the initial leakage at the sender will results in
comparable amount of leakage into the second excited state of qubit B (the receiver). 
Also, we plot the total population on doubly-excited qubit states  (normalized by the factor $\mathrm{er}$)
in Fig.~\ref{leak}(c). By comparing (b) and (c) we find that using stronger couplings (blue curves)
leads to less relative populations on qubit A (the sender). 
The population of photons left in the channel modes can be obtained from the
results of qubit populations by 
the normalization condition. We find that it is comparable to the leakage
of the qubits.

\subsection{Channel Loss}
\begin{figure}[t]
    \centering
    \includegraphics[width=0.8\textwidth]{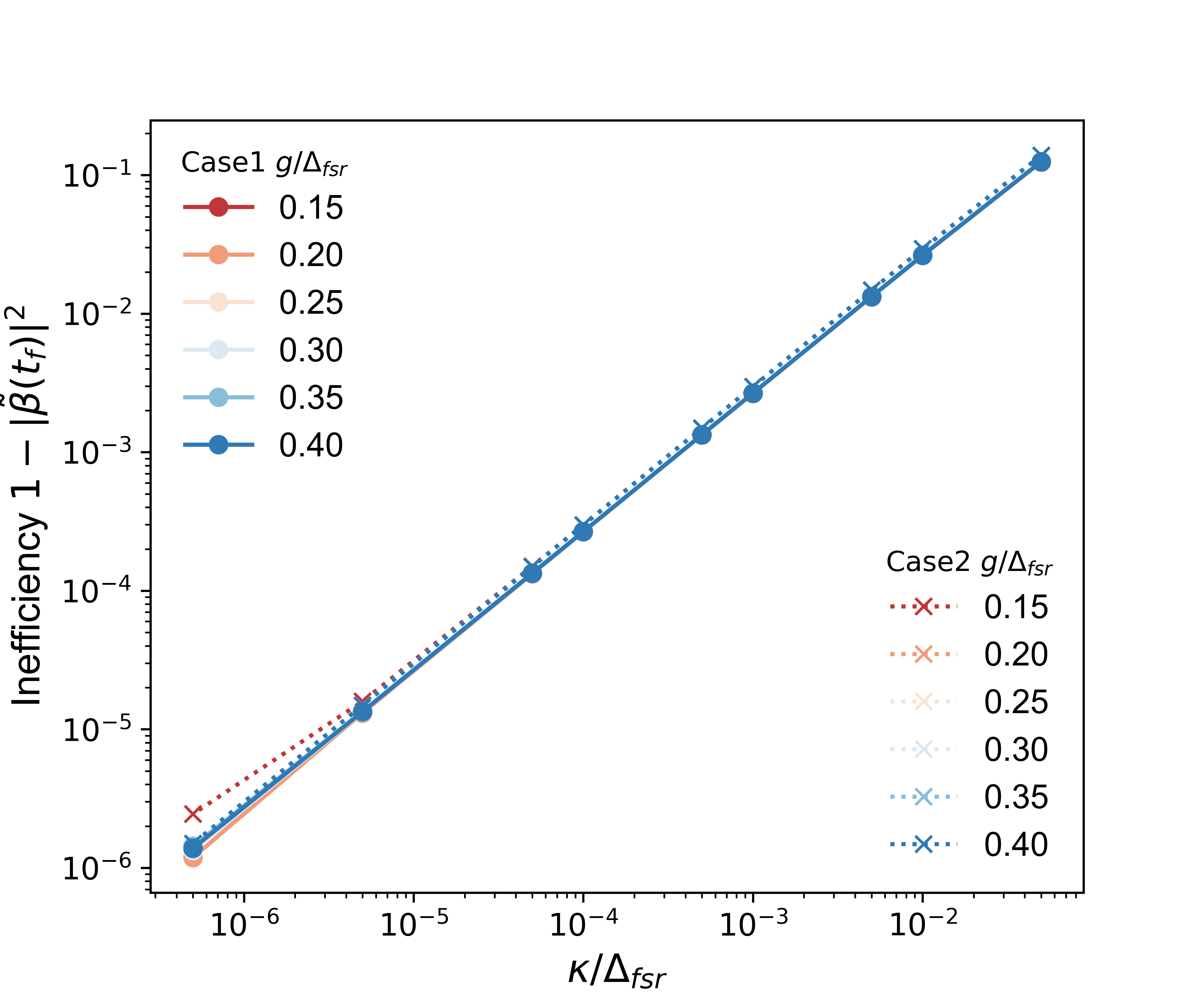} 
    \caption{Transfer inefficiency due to channel decoherence $\kappa$.} \label{channel_loss}
\end{figure}

When only the channel decoherence, 
with a rate $\kappa$, is considered, the inefficiency is found insensitive to the coupling strength 
as shown in Fig.{~}\ref{channel_loss}. This feature reflects our
analytical result that the integrated channel mode population $\mathcal{E}$ is almost invariant 
with respect to $g/\fsr$. 

\subsection{Qubit Loss}
\begin{figure}[h!]
    \centering
    \includegraphics[width=0.8\textwidth]{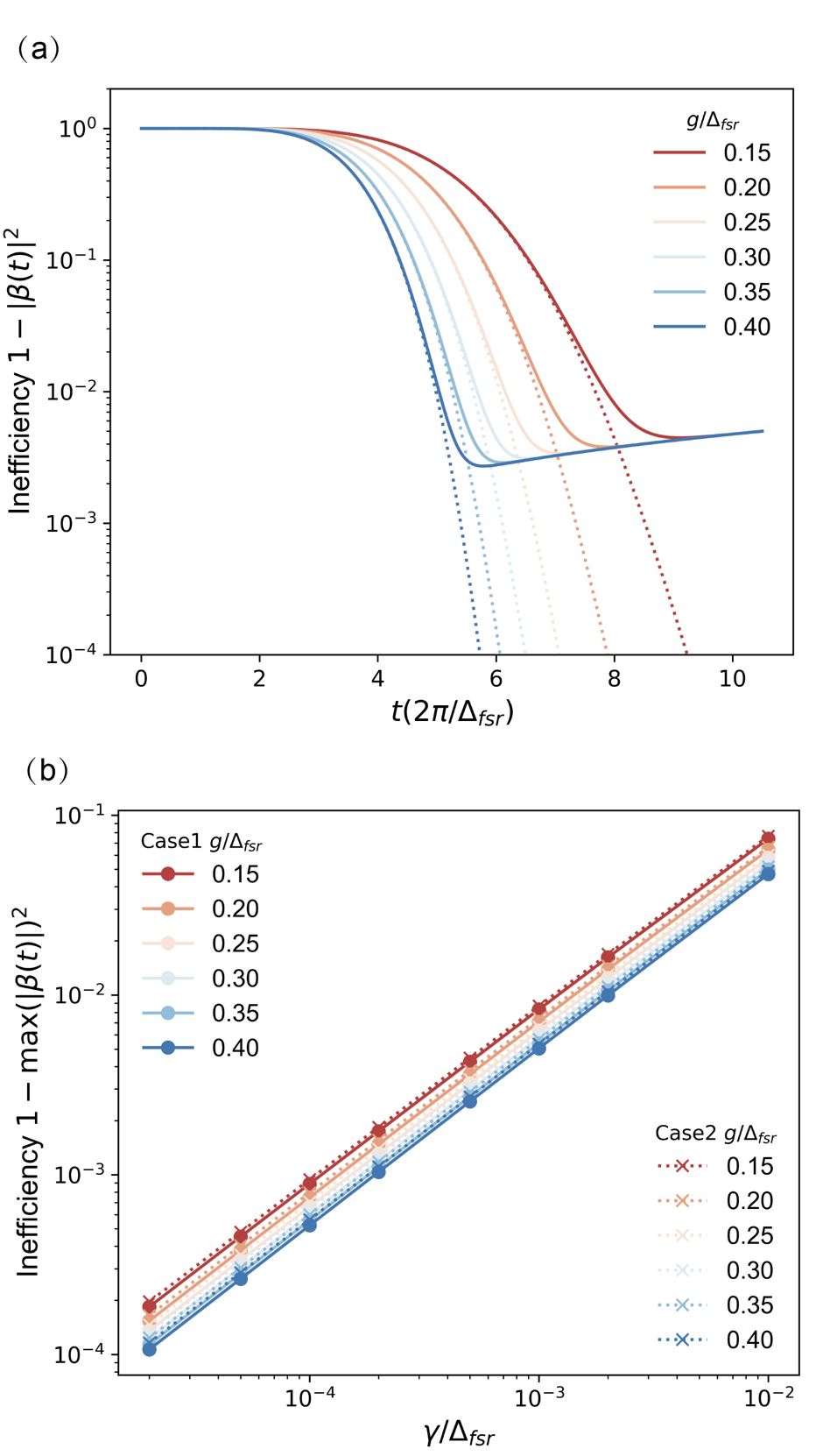} 
    \caption{Numerical results cosnidering qubit decoherence $\gamma$.
    (a) Dynamics at the receiving node. We consider Case 2 in the presence of $\gamma/\fsr=0.001$.
    (b) Transfer inefficiency due to qubit decoherence $\gamma$.} 
    \label{qubit_loss}
\end{figure}

Now we assume the channel is perfect and consider the influence of
qubit loss.  In this case, we have to assign a terminal point of the QST scheme. 
This is because when there considerable populations on qubit B,
it become more vulnerable to qubit decay. Thus, there must be a critical point where
the highest fidelity is reached.
We confirm this intuition by Fig.{~}\ref{qubit_loss}(a), which shows that the QST inefficiency 
goes up after reaching
some minimal inefficiency, due to the exponential 
decay of B [to compare, the dotted curves in Fig.~\ref{qubit_loss}(a) show the results
without qubit decay]. Therefore, we consider the transfer procedure accomplished when the 
inefficiency reaches its minimum, and plot the corresponding value in Fig{~}\ref{qubit_loss}(b). 
Without a surprise, larger $g$ results in a faster QST and therefore less qubit loss.

In our plots, we use the free spectral range $\Delta_{\text{fsr}}$ as the unit of frequency. If we assume that
$\Delta_{\text{fsr}}=2\pi \times 100\text{MHz}$, the decay rate of
$0.001 \Delta_{\text{fsr}}$ corresponds to a short qubit $T_1 \approx 1.6 \mu \text{s}$. 
Figures{~}\ref{channel_loss} and {~}\ref{qubit_loss} shows that even under such circumstance, 
our scheme can realize a rather small transfer inefficiency below 1\%.

\section{The Influence of Thermal Noise}\label{Sec:Thermal}

The channel is not in a vacuum state in realistic setups, 
especially considering that usually the
channel is not deeply contained inside a dilution refrigerator. For a rough estimation, we do not
specify the frequency-dependence of thermal populations but simply assign a
uniform distribution. Namely, we assume that every channel mode 
is initialized in a thermal Gibbs state that has the same average photon number 
denoted by $n_{\text{th}}$.

\begin{figure}[b]
    \centering
    \includegraphics[width=0.85\textwidth]{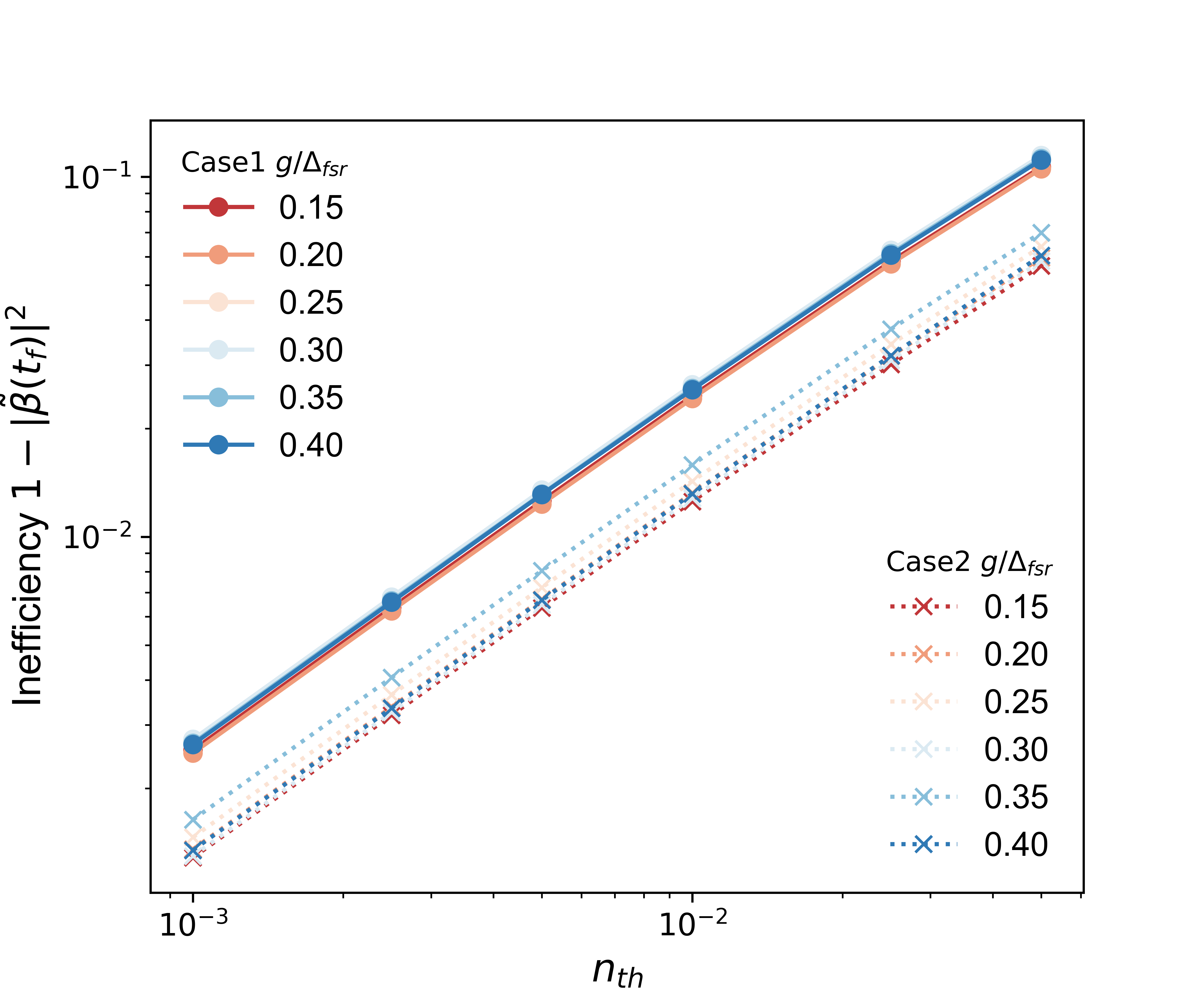} 
    \caption{Transfer inefficiency due to thermal noise in the channel. The channel
    initial state is assumed to be a thermal Gibbs state with a uniform 
    average photon number $n_{\text{th}}$ in each channel.}
    \label{thermal_channel}
\end{figure}

The results of QST inefficiency are plotted in Fig.{~}\ref{thermal_channel}.
For realistic superconducting circuits, residual photon number can reach 0.01 as stated 
in Ref.~\cite{Marquet:2024aa}. Then Fig.~\ref{thermal_channel} seems to show that thermal noise, 
among qubit leakage and decoherence, is the most devastating factor. 

\subsection{Qubits $\rightarrow$ Harmonic Resonators}

However, thermal noise can be tamed by using linear setups. 
It has been found in Refs.~\cite{Vermersch:2017aa,Xiang:2017aa}.
that the channel occupations, though affects the
evolution during QST, will not change the result of QST, provided that we replace the qubits with
harmonic resonators. 
We find the same conclusion in the multimode-resonator regime.
Explicitly, we use the same sets of control pulse $\{ g_A(t), g_B(t)\}$ but assume that
the qubit state has been locally written into the lowest two levels of a harmonic oscillator. Now the Hamiltonian reads
\begin{equation}
\begin{aligned}
    H(t)=&   g_A(t)\hat{a}^\dagger  \sum_k  \hat{c}_k e^{-i\delta_k t}  \\
 &\quad +g_B(t)\hat{b}^\dagger\sum_k (-1)^k\hat{c}_k e^{-i\delta_k t}+ \text{H.c.}
 \end{aligned}
\end{equation}
where $\hat{a}$ and $\hat{b}$ are bosonic annihilation operators of A and B, respectively. 
We assume the product initial state in the form of 
$\ket{1_A, 0_B}\bra{1_A, 0_B}\otimes\rho_{\text{ch}}$, where $\rho_{\text{ch}}$ denotes the channel state.
To demonstrate that our QST scheme works for any $\rho_{\text{ch}}$, it is much more convenient to
work in the Heisenberg picture of the above time-dependent Hamiltonian. Our target is to 
show that $\hat{b}(T)=\hat{a}(0)$, where 
we have assumed that QST starts at $t=0$ and ends at $t=T$.

The Heisenberg equation gives us the following equations of motion
\begin{equation}
    \dot{\hat{a}}(t) = -i g_A(t) \sum_k e^{-i\delta_k t} \hat{c}_k(t),
\end{equation}
\begin{equation}
    \dot{\hat{b}}(t) = -i g_B(t) \sum_k (-1)^k e^{-i\delta_k t} \hat{c}_k(t),
\end{equation}
where the channel mode annihilation operators can be formally written as
\begin{equation}
\begin{aligned}
    \hat{c}_k(t) = \hat{c}_k(0)   -i \int_{t_i=0}^{t} d\tau  &   [g_A^*(\tau)\hat{a}(\tau) \\
 &  + (-1)^k g_B^*(\tau) \hat{b}(\tau)]e^{i\delta_k \tau}.
  \end{aligned}
\end{equation}
These equations look similar to the Schr\"{o}dinger equation of the
single-excitation ansatz. But the difference
is that we cannot neglect $\hat{c}_k(0)$ as it is an operator. All the effects of the
thermal populations arise from the this term.

However, we notice that the linear bosonic dynamical system is ``closed'' in the sense that 
the Heisenberg picture operator $\hat{a}(t)$,
$\hat{b}(t)$, and every $\hat{c}_k(t)$ are linear combinations of $\hat{a}(0)$,
$\hat{b}(0)$ and $\{\hat{c}_k(0)\}_k$. This allows us to introduce the following expansions
\begin{equation}
\begin{aligned}
\hat{a}(t) &=\alpha(t)\hat{a}(0)+\alpha_b(t)\hat{b}(0)+\sum_{k}\alpha_k(t)\hat{c}_k(0), \\
\hat{b}(t) &=\beta(t)\hat{a}(0)+\beta_b(t)\hat{b}(0)+ \sum_{k}\beta_k(t)\hat{c}_k(0), \\
\hat{c}_k(t) & = c_k(t)\hat{a}(0)+ c_{k,b}(t)\hat{b}(0)+\sum_{j} \eta_{k,j}(t)\hat{c}_j(0).
\end{aligned}
\end{equation}
Therein, the coefficients have to satisfy certain normalization restrictions. For example, the 
basic bosonic commutation relation
$[\hat{b}(t), \hat{b}^\dagger (t)]=1$ implies that
\begin{equation}\label{norm_commu}
\abs{\beta(t)}^2+\abs{\beta_b(t)}^2+\sum_k\abs{\beta_k(t)}^2=1.
\end{equation}
Therefore, the Heisenberg equations can be reduced to the coupled equations for the 
coefficients of each operator at $t=0$. 

We focus on the
equations for the coefficients of $\hat{a}(0)$, i.e., $\alpha(t)$, $\beta(t)$ and $c_k(t)$.
These equations have exactly the same expression
with the single-excitation Schr\"{o}dinger equation.
Immediately, the success of QST in the single-excitation ansatz implies the
coefficients in the Heisenberg picture 
$\tbeta(t_f)\approx 1$. Then the normalization condition
Eq.~\eqref{norm_commu} directly gives the expected result that up to
exponentially corrections,
 $$\hat{b}(T)= \hat{a}(0).$$
It means that our QST scheme is immune to thermal channel population
after replacing qubits with harmonic oscillators.

\vspace{1.8 cm}

\bibliography{StateTransfer.bib}